\documentclass[]{llncs} 
\usepackage{a4} 
\usepackage{fleqn} 
\usepackage{lscape} 
\usepackage{amsfonts} 
\usepackage{caption2} 
\usepackage{amssymb} 
\usepackage{amstext} 
\usepackage{amsmath} 
\usepackage{pifont} 
\usepackage[bbgreekl]{mathbbol} 
\usepackage{stmaryrd} 
\usepackage{array}
\newcommand{\toprule}{\hline}
\newcommand{\bottomrule}{\hline}
\newcommand{\midrule}{\hline}

\usepackage[dvips]{epsfig} 
\usepackage[amsmath,thmmarks,thref]{ntheorem} 
\usepackage{alltt} 
\usepackage{ifthen} 
\usepackage{boxedminipage} 
\usepackage{graphicx} 
\usepackage[dvips]{color} 
\definecolor{svetlosivo}{gray}{0.85} 
\definecolor{sivo}{gray}{0.50}

\newlength{\blalength}%
\newenvironment{Entry}%
  {\begin{list}{}%
     {%
\setlength{\blalength}{\labelwidth}%
\addtolength{\blalength}{\labelsep}%
\setlength{\leftmargin}{\blalength}%
     }%
  }%
  {\end{list}}
      
\newsavebox{\blabox}
\newcommand{\Lentrylabel}[1]{%
  \sbox{\blabox}{#1}%
  \settowidth{\blalength}{\usebox\blabox}%
  \ifthenelse{\lengthtest{\blalength > \labelwidth}}
     {\parbox[b]{\labelwidth}
         {\makebox[0pt][l]{\usebox{\blabox}}\\}}%
     {\usebox{\blabox}}
  \hfil\relax}

\newif\ifProlog
\global\Prologfalse

\newenvironment{myv0}
{\begin{alltt}\Prologfalse\rm{}}%  
{\end{alltt}}

\newenvironment{myv}[1]%
{\setlength{\labelwidth}{10pt}%
\begin{Entry}%
\item[#1]%
\begin{alltt}\Prologtrue\cods}%  
{%
\end{alltt}%
\end{Entry}}

\theoremsymbol{}
\theorembodyfont{\upshape} 
\theoremstyle{fakethmstyle}

\newenvironment{mytable}
{\vbox\bgroup\begin{faketable*}[mytable]\unskip}%
{\unskip\end{faketable*}\unskip\egroup}

\newenvironment{mytable0}
{\begin{faketable*}[mytable0]\unskip}%
{\unskip\end{faketable*}\unskip}

\newenvironment{myfigurebox}[1]%
{%
\setlength{\fboxsep}{2mm}%
\edef\blablub{\noexpand#1}%
\begin{fakefigure*}[myfigurebox]%
\ifvmode\else\unskip\medskip\fi%
\begin{boxedminipage}[t]{\linewidth}%
}%
{%
\blablub\vspace*{-.5em}%
\end{boxedminipage}%
\end{fakefigure*}%
\bigskip%
}

\newcommand{\desno}{\ensuremath{\,\rightarrowtriangle\,}} 
\newcommand{\levo}{\ensuremath{\,\leftarrowtriangle\,}} 
\newcommand{\desnos}{\ensuremath{\stackrel{*}{\desno}}} 
\newcommand{\levos}{\ensuremath{\mathrel{\stackrel{*}{\levo}}}} 

\makeatletter
\let\ifempty\@ifempty 
\makeatother

\newcommand{\mathmeta}[1]{\ensuremath{\mathit{#1}}}

\newcommand{\meta}[1]{\mathmeta{#1}}

\newcommand{\mathobject}[1]{\ensuremath{\mathsf{#1}}}

\newcommand{\object}[1]{\mathobject{#1}}
\newcommand{\objects}{\sf}

\newcommand{\mathport}[1]{\mathmeta{#1}}

\newcommand{\mathgoal}[1]{\mathobject{#1}}

\newcommand{\mathanc}[1]{\mathobject{#1}}

\newcommand{\mathbet}[1]{\mathmeta{#1}}

\newcommand{\event}[4]{\ensuremath{\mathport{#1}\mathinner\mathgoal{#2}\ifempty{#3#4}{}{{\langle\textstyle\frac{\mathbet{#4}}{\mathanc{#3}}\rangle}}}}

\newcommand{\mybfbgr}{\boldsymbol{\{}}
\newcommand{\mybfegr}{\boldsymbol{\}}}
\newcommand{\mybfsep}{\boldsymbol{,\, }}
\newcommand{\mysep}{{,\, {}}}

\newcommand{\lineventblack}[4]{\ensuremath{\mathport{#1}\mathinner\mathgoal{#2}\ifempty{#3#4}{}%
{ \ifempty{#3}{}{\mybfsep\mybfbgr\mathanc{#3}\mybfegr}%
  \ifempty{#4}{}{\mybfsep\mybfbgr\mathbet{#4}\mybfegr} }}} 

\newcommand{\lineventshow}[4]{\ensuremath{\mathport{#1}\,\mathinner\mathgoal{#2}\ifempty{#3#4}{}%
{ \ifempty{#3}{}{\mysep\colorbox{svetlosivo}{\mathanc{#3}}}%
  \ifempty{#4}{}{\mysep\mathbet{#4}} }}}

\newcommand{\linevent}[4]{\ensuremath{\mathport{#1}\,\mathinner\mathgoal{#2}\ifempty{#3#4}{}%
{ \ifempty{#3}{}{\mysep\mathanc{#3}}%
  \ifempty{#4}{}{\mysep\mathbet{#4}} }}}

\newcommand{\cods}{\objects}

\newcommand{\true}{\object{true}}
\newcommand{\fail}{\object{fail}}
 
\newcommand{\prologneck}{\ensuremath{\mathrel{\mathord:\mathord-}}}
 
\newcommand{\prologif}{\ensuremath{\rightarrow}}

\newcommand{\myordeq}{\ensuremath{\mathord=}}

\newcommand{\type}[1]{\rm{\text{#1}}}
\newcommand{\typeopen}{\ensuremath{\boldsymbol{(}}}
\newcommand{\typeclose}{\ensuremath{\boldsymbol{)}}}
\newcommand{\typesep}{\ensuremath{\boldsymbol{,}}}
\newcommand{\typesei}{\ensuremath{\,::=\,}}
\newcommand{\hastype}{\ensuremath{\,:\,}}

\newcommand{\cons}{\ensuremath{\mathop{\bullet}}} 
\newcommand{\concat}{\ensuremath{\mathop{+}}}
\newcommand{\concatwo}{\ensuremath{\mathop{\ddagger}}}
\newcommand{\nil}{\meta{nil}} 

\newcommand{\gi}{\meta{A}} 
\newcommand{\gii}{\meta{B}} 
\newcommand{\giii}{\meta{C}} 
\newcommand{\query}{\meta{Q}} 
\newcommand{\goal}{\meta{G}} 
\newcommand{\body}{\meta{B}} 
\newcommand{\gai}{\ensuremath{\meta{G}_\meta{A}}} 
 
\newcommand{\head}{\meta{H}} 

\newcommand{\term}{\meta{T}}
\newcommand{\ti}{\ensuremath{\meta{T}_\meta{1}}} 
\newcommand{\tii}{\ensuremath{\meta{T}_\meta{2}}} 
\newcommand{\tagi}{\object{1}} 
\newcommand{\tagii}{\object{2}} 
\newcommand{\tagn}{\meta{N}} 
\newcommand{\ai}{\meta{U}} 
\newcommand{\bi}{\ensuremath{\mit\Sigma}} 

\newcommand{\progi}{\ensuremath{\mit\Pi}}
\newcommand{\queri}{\meta{Q}} 
\newcommand{\fsig}[2]{\ensuremath{\text{\object{#1}}/#2}} 

\newcommand{\substof}[1]{\ensuremath{\type{substOf}\typeopen#1\typeclose}}
\newcommand{\substofapl}[2]{\substapl{\substof{#1}}{#2}}
\newcommand{\substapl}[2]{\ensuremath{#1\typeopen#2\typeclose}}
\newcommand{\mguapl}[2]{\ensuremath{\type{mgu}\typeopen#1\typesep#2\typeclose}}

\newcommand{\Sel}[1]{\type{Sel(\meta{#1})}}
\newcommand{\mgui}{\ensuremath{\sigma}}

\newcommand{\substshow}[1]{\colorbox{sivo}{\bf\color{white}{\mathversion{bold}[#1]}}}

\newcommand{\bind}[2]{\ensuremath{#1\mathord/#2}}

\newcommand{\lema}[3]{\mathmeta{#1(\mathgoal{#2},\mathanc{#3})}}
\newcommand{\lemaatom}[2]{\lema{BY}{#1}{#2}}
\newcommand{\lemaor}[2]{\lema{OR}{#1}{#2}}

\newcommand{\astack}{A-stack}
\newcommand{\bstack}{B-stack}

\newcommand{\Astack}[1]{\ensuremath{\mathbb{#1}}}
\newcommand{\Bstack}[1]{\ensuremath{\mathbb{#1}}} 
\newcommand{\Ai}{\Astack{U}} 
\newcommand{\Aiz}{\Astack{U_0}} 
\newcommand{\Aii}{\Astack{V}} 

\newcommand{\Bi}{\Bstack{\Sigma}}
\newcommand{\Biz}{\Bstack{\Sigma_0}}
\newcommand{\Bio}{\Bstack{\Sigma_1}}

\newcommand{\Bdelta}[1]{\Bstack{\Delta_{#1}}}

\newcommand{\spectag}[1]{\tag{S:#1}}
\newcommand{\specref}[1]{\eqref{spec:#1}}
\newcommand{\speclabel}[1]{\label{spec:#1}\spectag{#1}}

\newcommand{\infsei}{\ensuremath{\,=\,}} 
\newcommand{\sei}{\ensuremath{\mathrel{\joinrel{:=}}}}

\newcommand{\idxp}[1]{\index{#1|primaryref}}
\newcommand{\idx}[1]{\index{#1}}

\newcommand{\idxn}[2]{\index{#1, #2}}

\newcommand{\mymodel}{S:PP} 
\newcommand{\wrt}{with respect to}

\newcommand{\ie}{i.\,e.\@}
\newcommand{\eg}{e.\,g.\@}

\begin{document}
\def\what{Pure Prolog Execution in 21 Rules} 
\title{\what}
\author{M. Kula\v{s}}
\institute{FernUniversit\"{a}t Hagen, FB Informatik, D-58084 Hagen, Germany\\
\email{marija.kulas@fernuni-hagen.de}
}
\maketitle           
\begin{abstract}
A simple mathematical definition of the 4-port model for pure Prolog is
given.  The model combines the intuition of ports with a compact
representation of execution state.  Forward and backward derivation
steps are possible.  The model satisfies a modularity claim, making it
suitable for formal reasoning.
\end{abstract}
\section{Introduction}
In order to 
formally handle 
(specify and prove) 
some properties of Prolog execution, 
we needed above all
a definition of a port.
A port is perhaps the single most popular notion
in Prolog debugging,
but theoretically 
it appears
still 
rather elusive.
The notion stems from the seminal article 
of L.\,Byrd \cite{byrdModel}
which identifies four different types of control flow in 
a Prolog execution,
as movements in and out of
procedure 
\emph{boxes}
via the four 
\emph{ports}
of these boxes:
\begin{itemize}
\item
\emph{call},
entering the procedure 
in order to solve a goal,
\item
\emph{exit},
leaving the procedure after a success, 
\ie\ a solution for the goal is found,
\item
\emph{fail},
leaving the procedure after the failure,
\ie\ there are no (more) solutions,
\item
\emph{redo},
re-entering the procedure,
\ie\ another solution is sought for.
\end{itemize}
In this work,
we present 
a formal definition
of ports,
which is 
a calculus of 
execution states, 
and 
hence 
provide
a formal model of
pure
Prolog execution,
\mymodel.
Our approach
is to define 
ports by virtue of their effect, as
\emph{port transitions}.
A port transition
relates two 
\emph{events}.
An event is
a state 
in the execution
of a given query 
\queri\ 
\wrt\ 
a given Prolog program \progi.
There are two restrictions we make:
\begin{enumerate}
\item
the program 
\progi\ 
has to be pure 
\item
the program 
\progi\ 
shall first be transformed into a canonical form.
\end{enumerate}
The first restriction
concerns only 
the presentation 
in this paper,
since 
our model
has 
been prototypically 
extended
to cover 
the control flow of full Standard Prolog,
as given in \cite{iso:deransart}.
The canonical form we use
is the common
single-clause representation.
This representation
is arguably
`near enough'
to the original program,
the only differences 
concern the head-unification
(which is now 
delegated to
the body)
and
the choices 
(which are now 
uniformly expressed 
as disjunction).

\section{Preliminaries and the main idea}

First we define the canonical form,
into which 
the 
original program 
has to 
be transformed.
Such a 
syntactic
form 
appears 
as an intermediate stage
in
defining 
the Clark's completion of a 
logic program,
and is
used 
in 
logic
program 
analysis. 
However, we are not aware of any 
consensus upon the
name for this form.
Some of the 
names in the literature
are
\emph{single-clausal form} \cite{lindgren:tr} and 
\emph{normalisation of a logic program} \cite{king}. 
Here we
use 
the name
\emph{canonical 
form},
partly on the grounds 
of our imposing a
transformation
on if-then as well
(this
additional
transformation
is
of no interest 
in the present paper, 
which has
to do only with pure Prolog,
but we state it for completeness).

\begin{mydefinition}[canonical form of a predicate]\label{def:canon} 
We say that a predicate \fsig{\meta{P}}{n}
is in the canonical form,
if its definition consists of a single clause
\(
P(X_1, ..., X_n) \prologneck B; Bs. 
\)
Here 
\meta{B}
is a "canonical body",  of the form
\(
X_1\mathord=T_1, \,\ldots, \,X_n\mathord=T_n, \,G, \,Gs
\), 
and
\(P(X_1, ..., X_n)\)
is a "canonical head", \ie\ 
\(X_1, ..., X_n\)
are distinct variables not appearing in 
\meta{G,Gs,T_1,...,T_n}. 
Further,
\meta{Bs} is a disjunction of canonical bodies  (possibly empty),
\meta{Gs} is a conjunction of 
goals 
(possibly empty),
and
\meta{G} is 
a goal
(for facts: 
\true).
Additionally, each if-then goal 
\meta{A\prologif B}
must be 
part of an if-then-else
(like \meta{A\prologif B; fail}).
\end{mydefinition}

\begin{myexample}[canonical form]
For the following program
\begin{myv}{}
q(a,b).
q(Z,c) \prologneck r(Z).
r(c).
\end{myv}
we obtain as canonical form
\begin{myv}{}
q(X,Y) \prologneck X=a, Y=b, true; X=Z, Y=c, r(Z).
r(X) \prologneck X=c, true. 
\end{myv}
{\vspace{-2em}$ $}
\end{myexample}
Having each predicate represented as one clause,
and bearing in mind the box metaphor 
above,
we identified
 some elementary execution steps.
For simplicity we first 
disregard variables.

\newcommand{\taga}{\ensuremath{\alpha}}
\newcommand{\tagb}{\ensuremath{\beta}}
The following table should 
give some intuition about 
the idea.
The symbols \taga, \tagb\ 
in this table
serve to 
identify
the appropriate redo-transition,
depending on the 
exit-transition.
Transitions are 
deterministic,
since the rules do not overlap.
\bigskip

\begin{mytable}\label{fig:port:intuit}
\begin{center}
\addtolength{\tabcolsep}{2pt}%
\setlength{\extrarowheight}{1pt}
\begin{tabular}{|c|llll|} 
\toprule
\multicolumn{1}{|c|}{\text{Term}} & \multicolumn{4}{l|}{\text{Port transitions in the context of Term}} 
\\ \midrule
\(\head\mathord:\mathord-\body\) & 
\(\event{call}{\head}{}{} \desno \event{call}{\body}{}{}\) & 
\(\event{exit}{\body}{}{} \desno \event{exit}{\head}{}{}\) & 
\(\event{fail}{\body}{}{} \desno \event{fail}{\head}{}{}\) &
\(\event{redo}{\head}{}{} \desno \event{redo}{\body}{}{}\) 

\\ 
\hline

\(\gi\mathord,\gii\) & 
\(\event{call}{{\gi\mathord,\gii}}{}{} \desno \event{call}{\gi}{}{}\) 
& \(\event{exit}{\gi}{}{} \desno \event{call}{\gii}{}{}\) 
& \(\event{fail}{\gi}{}{} \desno \event{fail}{{\gi\mathord,\gii}}{}{}\) 
& \(\event{redo}{{\gi\mathord,\gii}}{}{} \desno \event{redo}{\gii}{}{}\) \\

& &
\(\event{exit}{\gii}{}{} \desno \event{exit}{{\gi\mathord,\gii}}{}{}\) &
\(\event{fail}{\gii}{}{} \desno \event{redo}{\gi}{}{}\) &
 
\\ 
\hline

\(\gi\mathord;\gii\) & 
\(\event{call}{{\gi\mathord;\gii}}{}{} \desno \event{call}{\gi}{}{}\) & 
\(\event{exit}{\gi}{}{} \desno^\taga \event{exit}{{\gi\mathord;\gii}}{}{}\) &
\(\event{fail}{\gi}{}{} \desno \event{call}{\gii}{}{}\) &
\raisebox{.8ex}{\scriptsize \taga}\(\event{redo}{{\gi\mathord;\gii}}{}{} \desno \event{redo}{\gi}{}{}\) \\

& &
\(\event{exit}{\gii}{}{} \desno^\tagb \event{exit}{{\gi\mathord;\gii}}{}{}\) &
\(\event{fail}{\gii}{}{} \desno \event{fail}{{\gi\mathord;\gii}}{}{}\) &
\raisebox{.8ex}{\scriptsize \tagb}\(\event{redo}{{\gi\mathord;\gii}}{}{} \desno \event{redo}{\gii}{}{}\) 

\\ 
\hline

\(\object{true}\) &
\(\event{call}{\object{true}}{}{} \desno \event{exit}{\object{true}}{}{}\) & & &
\(\event{redo}{\object{true}}{}{} \desno \event{fail}{\object{true}}{}{}\) 
\\ 
\hline

\(\object{fail}\) &
\(\event{call}{\object{fail}}{}{} \desno \event{fail}{\object{fail}}{}{}\) & & &
\\ \bottomrule
\end{tabular}
\setlength{\extrarowheight}{0pt}
\caption{The idea of port transitions}
\end{center}
\end{mytable}%

\begin{myremark}[general goals]
Observe that
we
extend the notion of a port,
initially conceived for predicates,
 to 
\emph{general goals}. 
The shifting of 
attention
from predicates to goals
is the key 
idea
of
this
approach.
\end{myremark}

\begin{mynotation}[distinguishing meta-level from object-level]
In the following we show object-level 
terms (\ie\ actual Prolog terms)  in 
sans serif, like \true.
Meta-level terms 
(\ie\ anything else in the calculus)
will be shown in 
italics, like 
\meta{call},\bi,
or
in blackboard font, like 
\Ai,\Bi.
\end{mynotation}

Each transition
pertains to 
a certain 
context,
as indicated in \thref{fig:port:intuit}.
In the next step
towards 
the new definition of ports
we shall make this
dependency explicit,
by adding a parameter to 
each event. 

\begin{myexample}[good, bad and main]
Relative to 
the program
\begin{myv}{}
main \prologneck good, bad.
good.
\end{myv}
there 
are
the following execution steps for \object{main}:

\begin{tabbing}
\event{call}{\object{main}}{}{} \desno\ \\
\quad \event{call}{\object{(good,bad)}}{}{} \desno\ \\
\quad \quad \event{call}{\object{good}}{}{} \desno\ \\
\quad \quad \quad \event{call}{\true}{}{} \desno\ \\
\quad \quad \quad \event{exit}{\true}{}{} \desno\ \\
\quad \quad \event{exit}{\object{good}}{}{}\desno\  \\
\quad \quad \event{call}{\object{bad}}{}{} \desno\ \\
\quad \quad \event{fail}{\object{bad}}{}{} \desno\ \\
\quad \quad \event{redo}{\object{good}}{}{} \desno\ \\
\quad\quad \quad \event{redo}{\true}{}{} \desno\ \\
\quad\quad \quad \event{fail}{\true}{}{} \desno\ \\
\quad \quad \event{fail}{\object{good}}{}{} \desno\ \\
\quad \event{fail}{\object{(good,bad)}}{}{} \desno\ \\
\event{fail}{\object{main}}{}{} 
\end{tabbing}
The indentations 
should 
suggest the 
context
of the transitions,
which is 
not very satisfying,
since we
want 
our representation 
to be entirely symbolic,
and therefore visual aspects may not be part of the definition.
So we 
provide the context information 
within the calculus, 
by means of a \emph{stack of ancestors},
or 
\astack.
Hereby we define the
immediate ancestor (the parent) of a goal
to be
the context of the transition.
On some reflection, this is not enough.
In case of a redo of an atomary goal, like \event{redo}{\true}{}{} above,
we need to know 
how the 
goal 
was resolved,
in order to 
see
the remaining alternatives.
Since it is possible, in full Prolog, 
that a predicate definition changes between an exit and a redo,
simply accessing the program 
would not 
guarantee the retrieval of the 
definition
effective
at the time of call. 
For this reason we 
memorize, 
at an exit of an atomary goal,
the 
effectively \idx{logical update view}
used definition
(more about this on page \pageref{logupdate:more}). 
Also, on exit from a disjunction, some kind of memoing 
of the used disjunct 
is necessary.
So we
tried combining
the memoing 
(both kinds of 
memos:
used definitions and used disjuncts)
with the 
administration of 
variable bindings,
into one 
\emph{stack of bets}, or 
\bstack.
One 
claim
of 
this paper 
is 
that
an
\astack\ and 
a
\bstack\
are sufficient to
represent
the
execution of 
pure
Prolog.
As 
an
illustration
of the two-stack idea,
let us show
the above derivation
in complete detail.
In 
Appendix 
\ref{appendix:exa}
an example with variables
is given.
Each stack is enclosed in parentheses,  
\cons\
separates the elements,
and
\nil\ marks
the bottom 
of a stack.

\begin{myv0}{}
\linevent{call}{main}{\{\nil\}}{\{\nil\}}
 \desno \linevent{call}{(good,bad)}{\{main\cons{}\nil\}}{\{\nil\}}
 \desno \linevent{call}{good}{\{\tagi/good,bad\cons{}main\cons{}\nil\}}{\{\nil\}}
 \desno \linevent{call}{true}{\{good\cons{}\tagi/good,bad\cons{}main\cons{}\nil\}}{\{\nil\}}
 \desno \linevent{exit}{true}{\{good\cons{}\tagi/good,bad\cons{}main\cons{}\nil\}}{\{\nil\}}
 \desno \linevent{exit}{good}{\{\tagi/good,bad\cons{}main\cons{}\nil\}}{\{\lemaatom{true}{good}\cons{}\nil\}}
 \desno \linevent{call}{bad}{\{\tagii/good,bad\cons{}main\cons{}\nil\}}{\{\lemaatom{true}{good}\cons{}\nil\}}
 \desno \linevent{fail}{bad}{\{\tagii/good,bad\cons{}main\cons{}\nil\}}{\{\lemaatom{true}{good}\cons{}\nil\}}
 \desno \linevent{redo}{good}{\{\tagi/good,bad\cons{}main\cons{}\nil\}}{\{\lemaatom{true}{good}\cons{}\nil\}}
 \desno \linevent{redo}{true}{\{good\cons{}\tagi/good,bad\cons{}main\cons{}\nil\}}{\{\nil\}}
 \desno \linevent{fail}{true}{\{good\cons{}\tagi/good,bad\cons{}main\cons{}\nil\}}{\{\nil\}}
 \desno \linevent{fail}{good}{\{\tagi/good,bad\cons{}main\cons{}\nil\}}{\{\nil\}}
 \desno \linevent{fail}{(good,bad)}{\{main\cons{}\nil\}}{\{\nil\}}
 \desno \linevent{fail}{main}{\{\nil\}}{\{\nil\}}
\end{myv0}
{\vspace{-1em}$ $}
\end{myexample}

\section{The calculus \mymodel} 

We consider 
pure
Prolog
programs 
as 
given
in 
\thref{event:syntax},
syntax domain 
"\type{program}",
under restriction that every 
"\type{definition}"
has to be in the canonical form.

\begin{mydefinition}[event]\label{def:event} 
An \emph{event} is 
a quadruple 
\((\meta{Port}, \meta{Goal}, \meta{A\text{-}stack}, \meta{B\text{-}stack})\),
as 
given
by the grammar 
in 
\thref{event:syntax},
syntax domain "event".
\end{mydefinition}
Intuitively, an event is
a state of Prolog execution,
determined
by four parameters:
\begin{itemize}
\item
port
\item
current goal 
\item
history of
current 
goal
{(stack of 
generalized
ancestors, 
for short: \em\astack)}
\item
current 
environment
{(stack of 
generalized
bindings, \emph{bets}, 
for short: \em\bstack)}
\idxn{\astack}{stack of ancestors} 
\idxn{\bstack}{stack of bets} 
\end{itemize}
\begin{mydefinition}[transition rule]\label{def:transition} 
Let \progi\ be a program.
Port transition rules
wrt 
\progi\
are 
listed
in \thref{def:rules}. 
\end{mydefinition}

\begin{myfigurebox}{\caption{Language of events}\noexpand\label{event:syntax}}
\parindent .3cm

\begin{tabular}{lll}

\\[-.6em]
\bf\type{\bf event} &\typesei& \bf\large\event{\normalsize\type{port\,}}{\normalsize\type{goal\,}}{\type{stack of ancestors}}{\type{stack of bets}} \\[.4em] %
\bf\type{\bf event} &\typesei& \bf\lineventblack{\type{port\,}}{\type{goal}}{\type{stack of ancestors}}{\type{stack of bets}} \hfill\textmd{\% inline}\\[.4em]
\type{definition} & \typesei & \(\type{atom}\prologneck\type{goal}\) \\%
\type{program} & \typesei & \(\{\type{definition}.\}^{+}\) \\
\type{port} & \typesei & \meta{call}\ \textbar\ \meta{exit}\ \textbar\ \meta{fail}\ \textbar\ \meta{redo}\ \\
\type{goal}  & \typesei & \true\ \textbar\ \fail\ \textbar\ \type{atom} \textbar\ \type{term}=\type{term} \textbar\ \type{goal;goal} \textbar\ \type{goal,goal} \\ %
\type{ancestor} & \typesei & \true\ \textbar\ \fail\ \textbar\ \type{atom} \textbar\ \type{term}=\type{term} \textbar\ \type{tag/goal;goal} \textbar\ \type{tag/goal,goal} \\
\type{tag} & \typesei & \tagi\ \textbar\ \tagii\ \\
\type{memo} & \typesei & \lemaatom{\type{goal}}{\type{atom}} \textbar\ \lemaor{\type{goal}}{(\type{tag/goal;goal})} \\
\type{bet} & \typesei & \type{mgu} \textbar\ \type{memo} \\ %
\type{stack of Xs} & \typesei & \nil\ \textbar\ \type{X} \cons\ \type{stack of Xs} \\
\end{tabular}
\medskip

\noindent
Variables
\smallskip

\begin{tabular}{llllll}
\Ai, \Aii & \hastype & \type{stack of ancestors}, & \quad \ai & \hastype & \type{ancestor}  \\
\Bi, \Bdelta{}  & \hastype & \type{stack of bets}, & \quad \bi & \hastype & \type{bet} \\ 
\mgui & \hastype & \type{substitution}\\
\gi, \gii, \giii, \ \goal, \head  & \hastype & \type{goal} \\
\gai & \hastype & \type{atom} \\
\term & \hastype & \type{term} 
\end{tabular}
\medskip

\noindent
Semantic functions 
\smallskip

\begin{tabular}{lll}
\(\ti=\tii\) & \sei & \(\ti\) and \(\tii\) are identical \\ 
\substapl{\mgui}{\term} & \infsei & application of \mgui\ upon \term \\
\mguapl{\ti}{\tii} & \infsei & mgu of \ti\ and \tii \\
\substof{\Bi}  & \infsei & current substitution \infsei\ composition of all mgus from \Bi
\end{tabular}
\medskip 

\begin{tabular}{lll}
\substofapl{\nil}{\term} &\sei& \term \\
\substofapl{\bi\cons\Bi}{\term} &\sei& \(\left\{\begin{array}{ll} 
\substapl{\bi}{\substofapl{\Bi}{\term}},&\text{ if }\bi\text{ is an mgu} \\ 
\substofapl{\Bi}{\term},&\text{ if }\bi\text{ is a memo} \end{array}\right.\)
\end{tabular}
\medskip 

\noindent
Syntactic domains that we 
do not redefine, but take in their usual sense:
\noindent

\begin{tabular}{lll}
\type{term} (taken in the Prolog sense, as a superset of \type{goal}); \\
\type{atom} (atomary goal in logic programming); \\
\type{substitution, mgu}.
\end{tabular}
\idxn{\cons}{cons}%
\idxn{\emph{nil}}{empty stack}%
\end{myfigurebox}

\newpage
\thispagestyle{empty}
\def\deltawid{2cm}
\addtolength{\linewidth}{\deltawid} 
\newcounter{oldequation}%
\setcounter{oldequation}{\theequation}%
\renewcommand{\theequation}{spec:\arabic{equation}}%
\setcounter{equation}{0}%
\begin{myfigurebox}{\caption{Operational semantics \mymodel\ of pure Prolog}\noexpand\label{def:rules}} 
\addtolength{\mathindent}{-.3cm} 
\begin{align}
\nonumber\\[-1.2cm]
\intertext{Conjunction}
\event{call}{\gi,\gii}{\Ai}{\Bi} &\desno  \event{call}{\gi}{\tagi/\gi,\gii\cons\Ai}{\Bi} \speclabel{conj:1}\\ 
\event{exit}{\gi'}{\tagi/\gi,\gii\cons\Ai}{\Bi} &\desno  \event{call}{\gii''}{\tagii/\gi,\gii\cons\Ai}{\Bi}, \text{ with } \gii''\sei\substofapl{\Bi}{\gii} \speclabel{conj:2}\\ 
\event{fail}{\gi'}{\tagi/\gi,\gii\cons\Ai}{\Bi} &\desno  \event{fail}{\gi,\gii}{\Ai}{\Bi} \speclabel{conj:3}\\
\event{exit}{\gii'}{\tagii/\gi,\gii\cons\Ai}{\Bi} &\desno  \event{exit}{\gi,\gii}{\Ai}{\Bi} \speclabel{conj:4}\\ 
\event{fail}{\gii'}{\tagii/\gi,\gii\cons\Ai}{\Bi} &\desno  \event{redo}{\gi}{\tagi/\gi,\gii\cons\Ai}{\Bi} \speclabel{conj:5}\\
\event{redo}{\gi,\gii}{\Ai}{\Bi} &\desno  \event{redo}{\gii}{\tagii/\gi,\gii\cons\Ai}{\Bi} \speclabel{conj:6}
\intertext{Disjunction}
\event{call}{\gi;\gii}{\Ai}{\Bi} &\desno  \event{call}{\gi}{\tagi/\gi;\gii\cons\Ai}{\Bi} \speclabel{disj:1}\\
\event{fail}{\gi}{\tagi/\gi;\gii\cons\Ai}{\Bi} &\desno  \event{call}{\gii}{\tagii/\gi;\gii\cons\Ai}{\Bi} \speclabel{disj:2}\\
\event{fail}{\gii}{\tagii/\gi;\gii\cons\Ai}{\Bi} &\desno  \event{fail}{\gi;\gii}{\Ai}{\Bi} \speclabel{disj:3}\\
\event{exit}{\gi}{\tagi/\gi;\gii\cons\Ai}{\Bi} &\desno  \event{exit}{\gi;\gii}{\Ai}{\lemaor{\gi}{(\tagi/\gi;\gii)}\cons\Bi} \speclabel{disj:4}\\ 
\event{exit}{\gii}{\tagii/\gi;\gii\cons\Ai}{\Bi} &\desno  \event{exit}{\gi;\gii}{\Ai}{\lemaor{\gii}{(\tagii/\gi;\gii)}\cons\Bi} \speclabel{disj:5}\\ 
\event{redo}{\gi;\gii}{\Ai}{\lemaor{\giii}{(\tagn/\gi;\gii)}\cons\Bi} &\desno  \event{redo}{\giii}{\tagn/\gi;\gii\cons\Ai}{\Bi}\speclabel{disj:6}
\intertext{True}
\event{call}{\true}{\Ai}{\Bi} &\desno  \event{exit}{\true}{\Ai}{\Bi} \speclabel{true:1}\\
\event{redo}{\true}{\Ai}{\Bi} &\desno  \event{fail}{\true}{\Ai}{\Bi} \speclabel{true:2}
\intertext{Fail}
\event{call}{\fail}{\Ai}{\Bi} &\desno  \event{fail}{\fail}{\Ai}{\Bi} \speclabel{fail}
\intertext{Explicit unification}
\event{call}{\ti\mathord=\tii}{\Ai}{\Bi} &\desno  \begin{cases}
\event{exit}{\ti\mathord=\tii}{\Ai}{\mgui\cons\Bi}, & \text{if }\mguapl{\ti}{\tii}=\mgui \\
\event{fail}{\ti\mathord=\tii}{\Ai}{\Bi}, & \text{otherwise} 
\end{cases}\speclabel{unif:1}\medskip
\\
\event{redo}{\ti\mathord=\tii}{\Ai}{\mgui\cons\Bi} &\desno  \event{fail}{\ti\mathord=\tii}{\Ai}{\Bi}\speclabel{unif:2}
\intertext{User-defined atomary goal \gai}
\event{call}{\gai}{\Ai}{\Bi} &\desno  \begin{cases}
\event{call}{\substapl{\mgui}{\body}}{\gai\cons\Ai}{\Bi},
& \text{if } 
  \head\prologneck \body \text{ is a fresh renaming of a }\\
  &\hspace*{-2.7cm}\text{clause in \progi, }
      \text{and }\mguapl{\gai}{\head}=\mgui,\text{ and } \substapl{\mgui}{\gai}=\gai \\
\event{fail}{\gai}{\Ai}{\Bi}, & \text{otherwise} 
\end{cases}\speclabel{atom:1}\medskip%
\\
\event{exit}{\body}{\gai\cons\Ai}{\Bi} &\desno  \event{exit}{\gai}{\Ai}{\lemaatom{\body}{\gai}\cons\Bi} \speclabel{atom:2}\\ 
\event{fail}{\body}{\gai\cons\Ai}{\Bi} &\desno  \event{fail}{\gai}{\Ai}{\Bi} \speclabel{atom:3}\\
\event{redo}{\gai}{\Ai}{\lemaatom{\body}{\gai'}\cons\Bi} &\desno  \event{redo}{\body}{\gai'\cons\Ai}{\Bi} \speclabel{atom:4} 
\end{align}
\vspace*{-.5cm}
\end{myfigurebox}
\def\theequation{\arabic{equation}}%
\setcounter{equation}{\theoldequation}%
\addtolength{\linewidth}{-\deltawid} 

\subsection{Remarks on the calculus}
About event:
\begin{itemize}
\item
\emph{Current goal} 
is a generalization of \emph{selected literal}:
rather than focusing upon single literals, we focus upon goals.
\item
\emph{Ancestor} 
of a goal is 
defined in a 
disambiguating
manner, 
via
\emph{tags}.
\item
The notion of \emph{environment}
is generalized, to 
contain following \emph{bets}:
\begin{enumerate}
\item
variable bindings,
\item
choices taken (OR-branches),
\item
used predicate definitions.
\end{enumerate}
Environment is represented by one stack,
storing each bet
as soon as it is computed.
For an event to represent
the state of
pure Prolog execution, 
suffices
here
one environment and one ancestor stack.
\end{itemize}
About transitions:
\begin{itemize}
\item
Port transition relation is functional. The same holds for its converse, if restricted on \emph{legal events},
\ie\ events that can be reached from 
an \emph{initial event}
of the form
\event{call}{\goal}{\nil}{\nil}.
\item
This uniqueness of legal derivations enables \emph{forward and backward}
derivation steps,
in the spirit of the Byrd's article.
\item
\emph{Modularity} 
of derivation: The execution of a 
goal 
can be abstracted like 
for example
\(\event{call}{\goal}{\Ai}{\Bi}\desnos\event{exit}{\goal}{\Ai}{\Bdelta{}\concat\Bi}\).
Notice
the 
same 
\astack.
\end{itemize}

\begin{myremark}[atomary goal]
By \emph{atom} or \emph{atomary goal}
we denote  only user-defined predications.
So
\true, \fail\ or \(\ti\mathord=\tii\)
shall not be considered atoms.
\end{myremark}

\begin{myremark}[mgu]
The most general unifiers 
\mgui\
shall
be chosen to be idempotent,
\ie\
\(\substapl{\mgui}{\substapl{\mgui}{\term}} = \substapl{\mgui}{\term}\).
\end{myremark}

\begin{myremark}[tags]
The names \(\gi'\) or \(\gii'\) of 
\specref{conj:2}--\specref{conj:5}
should only suggest that the argument 
is
related to
\(\gi\) or \(\gii\),
but the actual retrieval is 
determined
by the 
tags
\tagi\ and \tagii,
saying that respectively the first or the second conjunct
are currently being 
tried.
For example, the rule 
\specref{conj:1}
states that the call of \(\gi,\gii\) leads to the call of \(\gi\) with immediate ancestor
\(\tagi/\gi,\gii\). 
This kind of 
add-on mechanism 
is necessary to be able to correctly handle 
a query like \meta{\gi,\gi} where retrieval by unification would 
get stuck on the first conjunct.
\end{myremark}

\begin{myremark}[canonical form]
Note the requirement 
\(\substapl{\mgui}{\gai}=\gai\)
in 
\specref{atom:1}.
Since the clauses are in canonical form,
unifying the head of a clause with a goal
could 
do no more than
rename the goal.
Since we do not need a renaming of the goal,
we 
may
fix 
the mgu to just 
operate on the clause.
\end{myremark}

\begin{myremark}[logical update view]
\label{logupdate:more}
Observe how 
\specref{atom:2}
and
\specref{atom:4}
serve
to 
implement
the  
\emph{logical update view}\idxp{logical update view}
of Lindholm and O'Keefe
\cite{logicalview},
saying
that the definition of a predicate
shall be fixed at the time of its  call.
This is further explained in the following remark.
\end{myremark}

\begin{myremark}["lazy" binding]
Although we memorize the used predicate definition \emph{on exit},
the 
definition will be 
unaffected by exit bindings,
because
\emph{bindings are applied lazily}:
Instead of 
"eagerly"
applying
any
bindings
as they occur
(\eg\ in \(\ti\mathord=\tii\), in resolution or in read),
we chose 
to do this only in conjunction
(in rule 
\specref{conj:2})
and nowhere else.
Due to the rules 
\specref{conj:1} and \specref{conj:4},
the exit bindings shall not affect
the predicate definition
like \eg\ \object{p(X) \prologneck q(X), r(X)}.

Also, lazy bindings 
enable
less 
`jumpy'
trace.
A 
jumpy
trace
can be illustrated by the following exit event (assuming we 
applied
 bindings
eagerly):
\[\linevent{exit}{\object{append([O],B,[O|B])}}{\{\tagii/([I|B]=[I|B]),append([],B,B)\cons\Ai\}}{\Bi}\] 
The 
problem 
consists in
exiting the goal \object{append([],B,B)} via \object{append([O],B,[O|B])},
the latter of course being no instance of the former.
By means of 
lazy binding, \idx{lazy binding}
we 
avoid the 
jumpiness,
and at the same time
make 
memoing 
definitions
on exit possible.
To ensure that the trace of a query execution
shows the 
correct
bindings,
an event
shall
be printed 
only after the current substitution
has been applied to it.

A perhaps more important
collateral %
advantage
of lazy binding 
is that 
a successful derivation
(see \thref{def:success})
can always 
be abstracted as follows:
\[\event{call}{\meta{Goal}}{}{} \desnos \event{exit}{\meta{Goal}}{}{}\]
even if \meta{Goal}
happened to 
get
further
instantiated 
in the course of 
this
derivation.
The instantiation will be 
reflected 
in the 
\bstack\
but not in the goal itself.
\end{myremark}

\section{Modelling Prolog execution}

\begin{mydefinition}[port transition relation, converse]
Let \progi\ be a program.
\emph{Port transition relation} 
\desno\
wrt \progi\
is 
defined 
in \thref{def:rules}. 
The
converse
relation
shall be denoted by \levo.
If \(E_1\desno E\),
we say that \(E_1\) \emph{leads to} \(E\).
An event \(E\) 
can be \emph{entered}, if
some event leads to it.
An event \(E\) 
can be \emph{left}, if
it leads to some event.
\end{mydefinition}

\begin{mylemma}
The relation \desno\ is 
functional,
\ie\ for each 
event \(E\) there can be at most one event \(E_1\)
such that \(E\desno E_1\).
\end{mylemma}
\begin{myproof}
The premisses of the
transition rules 
are mutually disjunct,
\ie\ there are no critical pairs.
\end{myproof}

\begin{myexample}[converse relation]\label{ex:rel:converse}
The converse of the port transition relation
is not functional,
since there may be more than one event leading to
the same event:
\begin{eqnarray*}
\event{call}{\ti\mathord=\tii}{\nil}{\nil} \desno \event{fail}{\ti\mathord=\tii}{\nil}{\nil}\\
\event{redo}{\ti\mathord=\tii}{\nil}{\mgui\cons\nil} \desno \event{fail}{\ti\mathord=\tii}{\nil}{\nil}
\end{eqnarray*}
We could have 
prevented
the
ambiguous situation
above
and
made converse relation functional as well,
by 
giving 
natural
conditions
on 
redo-transitions 
for 
atomary
goal and unification.
However, 
further down
it will be shown
that,
for events that are 
\emph{legal},
the converse relation is functional 
anyway.
\end{myexample}

\begin{mydefinition}[derivation]
Let \progi\ be a program.
Let \(E_0\), \(E\)
be 
events.
A \emph{\progi-derivation
of 
\(E\) from \(E_0\)},
written as
\(E_0 \desnos E\),
is a 
path from \(E_0\) to \(E\)
in 
the 
port transition 
relation
wrt \progi.
We say that \(E\) can be \emph{reached}
from \(E_0\).
\end{mydefinition}

\begin{mydefinition}[initial event, top-level goal\idx{goal!top-level goal, \emph{also} query}]
An \emph{initial event}
is any event of the form
\(\event{call}{\query}{\nil}{\nil}\),
where \query\ is a goal.
The goal \query\ 
of
an initial event
is called a
\emph{top-level goal},
or 
a \emph{query}.
\end{mydefinition}

\begin{mydefinition}[legal derivation, legal event, execution]
Let \progi\ be a program.
If there is a goal \query\
such that
\[ \event{call}{\query}{\nil}{\nil} \desnos E_0 \desnos E \]
is a \progi-derivation,
then we say that
\(E_0 \desnos E\)
is a \emph{legal \progi-derivation},
\(E\) is a \emph{legal \progi-event},
and
\( \event{call}{\query}{\nil}{\nil} \desnos E_0\)
is a 
\progi-\emph{execution} of 
the 
query
\query.
\end{mydefinition}

\begin{mydefinition}[final event]
A legal
event  
\(E\)
is a \emph{final} event
wrt program \progi,
if there is no
transition
\(E \desno E_1\)
wrt \progi.
\end{mydefinition}

\begin{mydefinition}[parent of goal\idx{goal!parent of a goal}]
If \(E= \event{Port}{\goal}{\Ai}{\Bi}\) is an event,
and \(\Ai = P\cons\Aii\),
then we say that \(P\) is the \emph{parent} of \goal.
\end{mydefinition}

\begin{mynotation}[selector tags\idxn{\emph{Sel}}{mapping a tagged parent to goal}] 
Function \Sel{\ai} 
is defined as follows:
\[\Sel{(\tagi/\gi,\gii)} \sei \gi,\ \Sel{(\tagii/\gi,\gii)} \sei \gii\]
and analogously
for disjunction.
\end{mynotation}

\begin{mydefinition}[push/pop event]
Let \(E\) be an event
with the port \(Port\).
If \meta{Port} is one of  
 \meta{call, redo},
then \(E\) 
is a \emph{push} event.
If \meta{Port} is one of  
\meta{exit, fail},
then \(E\) is a \emph{pop} event.
\end{mydefinition}

\begin{mylemma}[final event]\label{lem:finalevent}
If \(E\) is a legal pop event, and its \astack\ is not empty,
then 
\mbox{\(\exists E_1:\, E \desno E_1\)}
\end{mylemma}
\begin{myproof}[sketch]
According to the 
rules
(see also Appendix \ref{appendix:leave}),
the possibilities to 
leave
an exit event are:
\begin{align*}
\event{exit}{\gi'}{\tagi/\gi,\gii\cons\Ai}{\Bi} 
&\desno  \event{call}{\gii''}{\tagii/\gi,\gii\cons\Ai}{\Bi}, \text{ with } {\gii''}\sei\substofapl{\Bi}{\gii} \\ 
\event{exit}{\gii'}{\tagii/\gi,\gii\cons\Ai}{\Bi} 
&\desno  \event{exit}{\gi,\gii}{\Ai}{\Bi}\\ 
\event{exit}{\gi}{\tagi/\gi;\gii\cons\Ai}{\Bi} 
&\desno  \event{exit}{\gi;\gii}{\Ai}{\lemaor{\gi}{(\tagi/\gi;\gii)}\cons\Bi} \\
\event{exit}{\gii}{\tagii/\gi;\gii\cons\Ai}{\Bi} 
&\desno  \event{exit}{\gi;\gii}{\Ai}{\lemaor{\gii}{(\tagii/\gi;\gii)}\cons\Bi} \\
\event{exit}{\body}{\gai\cons\Ai}{\Bi} 
&\desno  \event{exit}{\gai}{\Ai}{\lemaatom{\body}{\gai}\cons\Bi}
\end{align*}
These rules 
state
that it is 
always
possible to 
leave
an
exit event \(\event{exit}{\goal}{\Ai}{\Bi}\), 
save for the following two restrictions:
The parent goal 
may not
be \true, \fail\ or a unification; 
and if the parent goal \meta{P} is a disjunction,
then there has to hold
\begin{equation}
\goal=\Sel{\meta{P}} \label{selprop}
\end{equation}
\ie\ it is not possible to leave an event
\(\event{exit}{\gi'}{\tagi/\gi;\gii\cons\Ai}{\Bi}\)
if \(\gi'\not=\gi\)
(and similarly for the second disjunct).
The first restriction is void,
since a parent cannot be \true, \fail\ or a unification anyway,
according to 
the 
rules.
It remains to show that the second restriction
is also void,
\ie\ a legal exit event
has necessarily the property \eqref{selprop}.
Looking at the rules for 
entering
an exit event,
we note that 
the goal part of an exit event 
either comes from the \astack,
or is 
\true\ or \(\ti\mathord=\tii\).
The latter two possibilities we may exclude, because
\(\event{exit}{\true}{\tagi/\gi;\gii\cons\Ai}{\Bi}\)
can only be derived from
\(\event{call}{\true}{\tagi/\gi;\gii\cons\Ai}{\Bi}\),
which cannot be reached
if \(\true\not=\gi\).
Similarly for unification.
So the goal part of a legal exit event 
must come from the \astack.
The elements of the \astack\
originate from call/redo events,
and they
have the property \eqref{selprop}.
In conclusion, 
we can always leave
a legal exit event with a nonempty \astack.
Similarly for a fail event.
\end{myproof}

\begin{myproposition}[uniqueness]\label{lem:uniq}
If \(E\) is a legal event,
then \(E\) 
can have 
only one legal predecessor,
and only one 
successor.
In case \(E\) is non-initial,  there is exactly one legal predecessor.
In case \(E\) is non-final,  there is exactly one 
successor.
\end{myproposition}
\begin{myproof}
The successor part follows from the functionality of \desno.
Looking at 
the 
rules,
we note that only two kinds of events may have 
more than one predecessor:
\event{fail}{\gai}{\Ai}{\Bi}
and
\event{fail}{\ti\mathord=\tii}{\Ai}{\Bi}.
Let \event{fail}{\ti\mathord=\tii}{\Ai}{\Bi} be a legal event.
Its predecessor may have been \event{call}{\ti\mathord=\tii}{\Ai}{\Bi},
on the condition that
\ti\ and \tii\ have no mgu (rule \specref{unif:1}),
or it could have been 
\event{redo}{\ti\mathord=\tii}{\Ai}{\mgui\cons\Bi}
(rule \specref{unif:2}).
In the latter case,
\event{redo}{\ti\mathord=\tii}{\Ai}{\mgui\cons\Bi}
must be a legal event,
so the \bstack\
\(\mgui\cons\Bi\)
had to be 
derived.
The only rule 
able to 
derive 
such a \bstack\
is \specref{unif:1},
on the condition that
the previous event was 
\event{call}{\ti\mathord=\tii}{\Ai}{\Bi}
and \(\mguapl{\ti}{\tii}=\mgui\).
Hence,
there can be only one legal
predecessor of \event{fail}{\ti\mathord=\tii}{\Ai}{\Bi},
depending solely on
\ti\ and \tii.
By a similar argument
we can prove that \event{fail}{\gai}{\Ai}{\Bi}
can have only one legal
predecessor.
This concludes the proof of functionality of the converse relation,
if restricted to the set of
legal events.
\end{myproof}

\begin{mynotation}[impossible event]
As a 
notational convenience,
all the
events which 
are not final 
and
do not lead to any further events 
by means of transitions 
\wrt\ 
the given program,
are 
said
to lead to the 
\emph{impossible event}, 
written as
\(\bot\).
Analogously for
events that 
are not initial events and
cannot be 
entered. 
In particular,
\( \event{redo}{\fail}{}{} \desno \bot\)
and
\( \event{exit}{\fail}{}{} \levo \bot\)
\wrt\ 
any 
program.
Some impossible events are: 
\event{call}{\goal}{\nil}{\mgui\cons\nil},
\event{redo}{\goal}{\nil}{\Bi} ({cannot be entered, non-initial}), 
and
\event{redo}{\object{p}}{\Ai}{\nil} ({cannot be left, non-final}).
\end{mynotation}

\begin{mylemma}[non-legal event]\label{lem:illegal}
If \(E\desnos \bot\), then \(E\) is not legal.
If
\(E\levos \bot\), then \(E\) is not legal.
\end{mylemma}
\begin{myproof}
Let \(E \levo E_1\). 
If \(E\) is legal,
then,
because of the uniqueness of the transition,
\(E_1\) has to be legal as well.
\end{myproof}

\begin{mylemma}[call is up-to-date]\label{lem:uptodate}
For a legal call event
\(
\event{call}{\goal}{\Ai}{\Bi}
\)
holds that \(\goal = \substofapl{\Bi}{\goal}\),
meaning that
the substitutions%
\
from the \bstack\ 
are 
\emph{already applied} 
upon the 
goal
to be called.
In other words, the goal of any 
legal
call event is 
up-to-date 
relative to
the current substitution.
\end{mylemma}
Notice that this property 
holds only for call events.

\begin{mynotation}[stack concatenation]
Concatenation of stacks 
we denote by \concat.
Concatenating
to
both 
stacks
of an event
we denote by \concatwo:
If \(E = \event{Port}{\goal}{\Ai}{\Bi}\), then
\(\event{E}{\concatwo}{\Aii}{\Bdelta} \sei 
\event{Port}{\goal}{\Ai\concat\Aii}{\Bi\concat\Bdelta}\).
\idxn{\concat}{concatenation}%
\idxn{\concatwo}{concatenation to both stacks}%
\end{mynotation}

\begin{myproposition}[modularity of derivation]
Let \progi\ be a program.
Let 
\meta{Pop} be one of \(\meta{exit}, \meta{fail}\).
If 
\[\event{call}{\goal}{\nil}{\nil} \desno E_1\desno...\desno E_n\desno \event{Pop}{\goal}{\nil}{\Bdelta{}}\]
is a 
legal \progi-derivation,
then
for every 
\astack\
\Ai\
and for every 
\bstack\
\Bi\
such that
\event{call}{\goal}{\Ai}{\Bi}
is a legal event,
holds:
\[\event{call}{\goal}{\Ai}{\Bi} \desno \event{\ensuremath{E_1}}{\concatwo}{\Ai}{\Bi}\desno...\desno \event{\ensuremath{E_n}}{\concatwo}{\Ai}{\Bi}\desno \event{Pop}{\goal}{\Ai}{\Bdelta{}\concat\Bi} \]
is also a 
legal %
\progi-derivation.
\end{myproposition}

\begin{myproof}
Observe that our rules (with the exception of \specref{conj:2})
refer only to 
the existence of the top element of 
some stack,
never to the 
emptiness
of a stack.
Since the 
top element of a 
stack \(S\) 
cannot
change after 
appending 
another stack 
to \(S\),
it is possible to 
emulate
each 
of the 
original
derivation steps
using
the 
`new'
stacks.

It remains to consider
the rule \specref{conj:2},
which applies the whole current substitution upon the second conjunct.
First note that any variables in a 
legal
derivation stem either from the 
top-level goal 
or are fresh. 
According to
the \thref{lem:uptodate},
a call event is always up-to-date,
\ie\ the current substitution has already been applied to the goal.
The most general unifiers 
may be chosen 
to be idempotent, so
a multiple application of a substitution
amounts to a single application.
Hence, if
\event{call}{\goal}{\Ai}{\Bi}
is a legal event,
the substitution of \Bi\
cannot affect any variables of the
original derivation.
\end{myproof}

\section{Applications}

\subsection{Specifying program properties} 
Uniqueness and modularity of legal port derivations
allow us to
succinctly define
some traditional notions.

\begin{mydefinition}[termination, success, failure]\label{def:success}
A goal \goal\
is said to terminate wrt program \progi,
if there is a \progi-derivation
\[\event{call}{\goal}{\nil}{\nil}\desnos \event{Pop}{\goal}{\nil}{\Bdelta{}}\]
where \meta{Pop} is one of \(\meta{exit},\meta{fail}\).
In case of \meta{exit}, 
the derivation is \emph{successful},
otherwise it is \emph{failed}.
\end{mydefinition}
In a failed derivation, \(\Bdelta{}=\nil\).

\begin{mydefinition}[computed answer]
In a successful derivation
\[\event{call}{\goal}{\nil}{\nil}\desnos \event{exit}{\goal}{\nil}{\Bdelta{}}\]
is \(\substof{\Bdelta{}}\),
restricted upon the variables of \goal,
called
the \emph{computed answer substitution} for \goal.
\end{mydefinition}

\subsection{Proving program properties}

Uniqueness of legal derivation steps
enables 
\emph{forward and backward}
derivation steps,
in the spirit of the 
Byrd's article.
Push events (call, redo) are 
more
amenable to 
forward steps,
and pop events (exit, fail)
are more amenable  
to backward steps.
We illustrate this 
by a small example.

\begin{mylemma}
If the events on the left-hand sides are legal,
the following are legal derivations
(for 
appropriate 
\Biz, \Bio): 
\begin{eqnarray}
\event{exit}{\gi;\gii,\fail}{\Ai}{\Bi}  &\levo & \event{exit}{\gi}{\tagi/\gi;\gii,\fail\cons\Ai}{\Biz} \label{constr:prop:ef}\\
\event{redo}{\gi;\gii,\fail}{\Ai}{\Bi}  &\desno & \event{redo}{\gi}{\tagi/\gi;\gii,\fail\cons\Ai}{\Bio} \label{constr:prop:rf}
\end{eqnarray}%
\end{mylemma}
\vspace*{-1em}
\begin{myproof}
The 
first statement
claims:
If \event{exit}{\gi;\gii,\fail}{\Ai}{\Bi} 
is legal,
then 
it 
was 
reached
via 
\event{exit}{\gi}{}{}. 
Without 
inspecting
\Bi,
in general 
it 
is 
not 
known
whether 
a disjunction
succeeded 
via
its first, or 
via
its second member.
But 
in this particular disjunction,
the second member cannot 
succeed: 
Assume 
there are some \Aiz, \Biz\ with
\(
\event{exit}{\gi;\gii,\fail}{\Ai}{\Bi} \levo \event{exit}{\gii,\fail}{\Aiz}{\Biz}
\).
According to the rules:
\[\event{exit}{\gii,\fail}{\Aiz}{\Biz} 
\levo \event{exit}{\fail}{\tagii/\gii,\fail\cons\Aiz}{\Biz}
\levo  \bot\]
So according to \thref{lem:illegal},
\event{exit}{\gii,\fail}{\Aiz}{\Biz}
is not a legal event, which
proves \eqref{constr:prop:ef}.
Similarly, the non-legal derivation
\text{\(
\event{redo}{\gii, \fail}{}{}  \desno \event{redo}{\fail}{}{}  \desno \bot 
\)}
proves 
\eqref{constr:prop:rf}.
\end{myproof}

\medskip
Modularity 
of legal derivations
enables 
\emph{abstracting
the execution} of a goal,
like in the following 
example.

\begin{myexample}[modularity]
Assume that a goal \gi\ succeeds,
\ie\ 
\(\event{call}{\gi}{\nil}{\nil}\desnos\event{exit}{\gi}{\nil}{\Bdelta{}}\).
Then we have the following legal derivation:
\begin{align*}
\event{call}{\gi,\gii}{\nil}{\nil} &\desno \event{call}{\gi}{\tagi/\gi,\gii\cons\nil}{\nil}, \text{ by \specref{conj:1}}\\
&\desnos \event{exit}{\gi}{\tagi/\gi,\gii\cons\nil}{\Bdelta{}\cons\nil}, \text{ by modularity and success of \gi}\\
&\desno \event{call}{\gii'}{\tagii/\gi,\gii\cons\nil}{\Bdelta{}\cons\nil},  \text{ by \specref{conj:2}}, \text{ where }\gii'=\substofapl{\Bdelta{}}{\gii}
\end{align*}
If \gi\ fails, then we have:
\begin{align*}
\event{call}{\gi,\gii}{\nil}{\nil} &\desno \event{call}{\gi}{\tagi/\gi,\gii\cons\nil}{\nil}, \text{ by \specref{conj:1}}\\
&\desnos \event{fail}{\gi}{\tagi/\gi,\gii\cons\nil}{\nil}, \text{ by modularity and failure of \gi}\\
&\desno \event{fail}{\gi,\gii}{\nil}{\nil}, \text{ by \specref{conj:3}}
\end{align*}
\end{myexample}

\section{Conclusions and outlook}
In this paper we 
give
a
simple mathematical definition
\mymodel\
of 
the 4-port model 
of
pure Prolog.
Some 
potential
for
formal verification of pure Prolog
has
been 
outlined.
There are two interesting
directions for 
future work 
in this area:

\medskip
\noindent
(1)
formal specification
of the control flow of \emph{full Standard Prolog}
(%
currently 
we have 
a
 prototype
for this, 
within 
the 4-port model)

\medskip
\noindent
(2)
formal
specification 
and proof
of
some
non-trivial program properties,
like adequacy and non-interference of a  practical program transformation.

\section{Related work}
Concerning
attempts to 
formally define
the 4-port model,
we are aware of 
only few
previous works.
One is
a graph-based model of Tobermann and Beckstein 
\cite{tobermann}, 
who
formalize
the 
graph traversal idea of Byrd,
defining the notion of a 
\emph{trace}
(of a given query \wrt\ a given program),
as
a path
in 
a 
trace graph.
The ports 
are 
quite
lucidly 
defined
as 
hierarchical 
nodes
of such a graph.
However, even for a simple recursive program and a ground query,
with a finite SLD-tree,
the corresponding trace graph is infinite,
which limits its applicability.
Another model of Byrd box is a continuation-based 
approach
of Jahier, 
Ducass\'{e}
and Ridoux \cite{jahier:ducasse:ridoux:00}.
There is also a stack-based attempt 
in \cite{KulasM:rewps},
but although it provides for some parametrizing,
it suffers essentially the same problem
as 
the 
continuation-based
approach,
and also
the prototypical implementation of
the tracer given in \cite{byrdModel},
taken as a specification
of Prolog execution:
In these three attempts,
a port is represented by
some 
semantic 
action (\eg\ writing of a message),
instead of a formal method.
Therefore it
is not clear
how to use any of these models 
to prove some port-related
assertions.

In contrast to the 
few
 specifications of the Byrd box,
there are 
many more
general models
of 
pure (or even full) Prolog
execution.
Due to space limitations
we mention here 
only
some 
models,
directly relevant 
to 
\mymodel,
and
for a more 
comprehensive discussion 
see 
\eg\ 
\cite{KulasBeierle:defsp}.
Comparable to 
our
work 
are
the stack-based approaches.
St\"ark gives  in 
\cite{staerk97full},
as a side issue,
a
simple operational semantics of pure logic programming.
A state of 
execution
is a 
stack
of frame stacks,
where each frame consists of a goal (ancestor) 
and an environment.
In comparison, 
our state of 
execution
consists
of exactly
{one environment and one ancestor stack}. 
The seminal paper of Jones and Mycroft \cite{jones:mycroft:84}
was the first to present
a stack-based model
of execution,
applicable
to pure Prolog with cut added.
It 
uses a sequence of frames. 
In 
these
stack-based approaches
(including our 
previous attempt 
\cite{KulasBeierle:defsp}), 
there is no 
\emph{modularity}, 
\ie\
it is not possible to abstract
the execution of a subgoal. 

\section*{Acknowledgments}
Many thanks for helpful comments are due to anonymous referees.

\begin{appendix}
\section{Leaving events}\label{appendix:leave} 
\def\deltawid{2cm}
\addtolength{\linewidth}{\deltawid} 
\setcounter{oldequation}{\theequation}%
\renewcommand{\theequation}{spec:\arabic{equation}}%
\setcounter{equation}{0}%
\begin{myfigurebox}{} 
\addtolength{\mathindent}{-.3cm} 
\begin{align}
\nonumber\\[-1.2cm]
\intertext{Leaving a call event}
\event{call}{\gi,\gii}{\Ai}{\Bi} &\desno  \event{call}{\gi}{\tagi/\gi,\gii\cons\Ai}{\Bi} \spectag{conj:1}\\ 
\event{call}{\gi;\gii}{\Ai}{\Bi} &\desno  \event{call}{\gi}{\tagi/\gi;\gii\cons\Ai}{\Bi} \spectag{disj:1}\\
\event{call}{\true}{\Ai}{\Bi} &\desno  \event{exit}{\true}{\Ai}{\Bi} \spectag{true:1}\\
\event{call}{\fail}{\Ai}{\Bi} &\desno  \event{fail}{\fail}{\Ai}{\Bi} \spectag{fail}\\
\event{call}{\ti\mathord=\tii}{\Ai}{\Bi} &\desno  \begin{cases}
\event{exit}{\ti\mathord=\tii}{\Ai}{\mgui\cons\Bi}, & \text{if }\mguapl{\ti}{\tii}=\mgui \\
\event{fail}{\ti\mathord=\tii}{\Ai}{\Bi}, & \text{otherwise} 
\end{cases}\spectag{unif:1}\medskip
\\
\event{call}{\gai}{\Ai}{\Bi} &\desno  \begin{cases}
\event{call}{\substapl{\mgui}{\body}}{\gai\cons\Ai}{\Bi},
& \text{if } 
  \head\prologneck \body \text{ is a fresh renaming of a }\\
  &\hspace*{-2.7cm}\text{clause in \progi, }
      \text{and }\mguapl{\gai}{\head}=\mgui,\text{ and } \substapl{\mgui}{\gai}=\gai \\
\event{fail}{\gai}{\Ai}{\Bi}, & \text{otherwise} 
\end{cases}\spectag{atom:1}\medskip%
\\
\intertext{Leaving a redo event}
\event{redo}{\gi,\gii}{\Ai}{\Bi} &\desno  \event{redo}{\gii}{\tagii/\gi,\gii\cons\Ai}{\Bi} \spectag{conj:6}\\
\event{redo}{\gi;\gii}{\Ai}{\lemaor{\giii}{(\tagn/\gi;\gii)}\cons\Bi} &\desno  \event{redo}{\giii}{\tagn/\gi;\gii\cons\Ai}{\Bi}\spectag{disj:6}\\
\event{redo}{\true}{\Ai}{\Bi} &\desno  \event{fail}{\true}{\Ai}{\Bi} \spectag{true:2}\\
\event{redo}{\ti\mathord=\tii}{\Ai}{\mgui\cons\Bi} &\desno  \event{fail}{\ti\mathord=\tii}{\Ai}{\Bi}\spectag{unif:2}\\
\event{redo}{\gai}{\Ai}{\lemaatom{\body}{\gai'}\cons\Bi} &\desno  \event{redo}{\body}{\gai'\cons\Ai}{\Bi} \spectag{atom:4}\\ 
\intertext{Leaving an exit event}
\event{exit}{\gi'}{\tagi/\gi,\gii\cons\Ai}{\Bi} &\desno  \event{call}{\gii''}{\tagii/\gi,\gii\cons\Ai}{\Bi}, \text{ with } \gii''\sei\substofapl{\Bi}{\gii} \spectag{conj:2}\\ 
\event{exit}{\gii'}{\tagii/\gi,\gii\cons\Ai}{\Bi} &\desno  \event{exit}{\gi,\gii}{\Ai}{\Bi} \spectag{conj:4}\\ 
\event{exit}{\gi}{\tagi/\gi;\gii\cons\Ai}{\Bi} &\desno  \event{exit}{\gi;\gii}{\Ai}{\lemaor{\gi}{(\tagi/\gi;\gii)}\cons\Bi} \spectag{disj:4}\\ 
\event{exit}{\gii}{\tagii/\gi;\gii\cons\Ai}{\Bi} &\desno  \event{exit}{\gi;\gii}{\Ai}{\lemaor{\gii}{(\tagii/\gi;\gii)}\cons\Bi} \spectag{disj:5}\\ 
\event{exit}{\body}{\gai\cons\Ai}{\Bi} &\desno  \event{exit}{\gai}{\Ai}{\lemaatom{\body}{\gai}\cons\Bi} \spectag{atom:2}\\ 
\intertext{Leaving a fail event}
\event{fail}{\gi'}{\tagi/\gi,\gii\cons\Ai}{\Bi} &\desno  \event{fail}{\gi,\gii}{\Ai}{\Bi} \spectag{conj:3}\\
\event{fail}{\gii'}{\tagii/\gi,\gii\cons\Ai}{\Bi} &\desno  \event{redo}{\gi}{\tagi/\gi,\gii\cons\Ai}{\Bi} \spectag{conj:5}\\
\event{fail}{\gi}{\tagi/\gi;\gii\cons\Ai}{\Bi} &\desno  \event{call}{\gii}{\tagii/\gi;\gii\cons\Ai}{\Bi} \spectag{disj:2}\\
\event{fail}{\gii}{\tagii/\gi;\gii\cons\Ai}{\Bi} &\desno  \event{fail}{\gi;\gii}{\Ai}{\Bi} \spectag{disj:3}\\
\event{fail}{\body}{\gai\cons\Ai}{\Bi} &\desno  \event{fail}{\gai}{\Ai}{\Bi} \spectag{atom:3} 
\end{align}
\vspace*{-.5cm}
\end{myfigurebox}
\def\theequation{\arabic{equation}}%
\setcounter{equation}{\theoldequation}%
\addtolength{\linewidth}{-\deltawid} 
%
\def\deltawid{1.5cm}
\addtolength{\textwidth}{\deltawid} 
\begin{landscape}
\section{An example with variables}\label{appendix:exa} 
\noindent
Assume the following program \progi:
\begin{myv}{}
post(X,Y) \prologneck one(X,Y), two(X,Y).
one(X,\_) \prologneck X=1.
two(\_,Y) \prologneck Y=a; Y=b.
\end{myv}
\thref{fig:exec} below shows the complete \progi-execution of the goal
\object{post(X,Y),fail} in the model \mymodel.  Highlighted are the
\astack s and the mgus.  Notice the "lazy" binding of variables in the
current goal.

\bigskip

\begin{mytable0}\label{fig:exec}
\begin{myv0}{}\scriptsize{}
\lineventshow{call}{(post(X,Y),fail)}{\{\nil\}}{\{\nil\}}
 \desno \lineventshow{call}{post(X,Y)}{\{1/post(X,Y),fail\cons{}\nil\}}{\{\nil\}}
 \desno \lineventshow{call}{(one(X,Y),two(X,Y))}{\{post(X,Y)\cons{}1/post(X,Y),fail\cons{}\nil\}}{\{\nil\}}
 \desno \lineventshow{call}{one(X,Y)}{\{1/one(X,Y),two(X,Y)\cons{}post(X,Y)\cons{}1/post(X,Y),fail\cons{}\nil\}}{\{\nil\}}
 \desno \lineventshow{call}{X\myordeq{}1}{\{one(X,Y)\cons{}1/one(X,Y),two(X,Y)\cons{}post(X,Y)\cons{}1/post(X,Y),fail\cons{}\nil\}}{\{\nil\}}
 \desno \lineventshow{exit}{X\myordeq{}1}{\{one(X,Y)\cons{}1/one(X,Y),two(X,Y)\cons{}post(X,Y)\cons{}1/post(X,Y),fail\cons{}\nil\}}{\{\substshow{\bind{X}{1}}\cons{}\nil\}}
 \desno \lineventshow{exit}{one(X,Y)}{\{1/one(X,Y),two(X,Y)\cons{}post(X,Y)\cons{}1/post(X,Y),fail\cons{}\nil\}}{\{\lemaatom{X\myordeq{}1}{one(X,Y)}\cons{}\substshow{\bind{X}{1}}\cons{}\nil\}}
 \desno \lineventshow{call}{two(1,Y)}{\{2/one(X,Y),two(X,Y)\cons{}post(X,Y)\cons{}1/post(X,Y),fail\cons{}\nil\}}{\{\lemaatom{X\myordeq{}1}{one(X,Y)}\cons{}\substshow{\bind{X}{1}}\cons{}\nil\}}
 \desno \lineventshow{call}{(Y\myordeq{}a;Y\myordeq{}b)}{\{two(1,Y)\cons{}2/one(X,Y),two(X,Y)\cons{}post(X,Y)\cons{}1/post(X,Y),fail\cons{}\nil\}}{\{\lemaatom{X\myordeq{}1}{one(X,Y)}\cons{}\substshow{\bind{X}{1}}\cons{}\nil\}}
 \desno \lineventshow{call}{Y\myordeq{}a}{\{(1/(Y\myordeq{}a);Y\myordeq{}b)\cons{}two(1,Y)\cons{}2/one(X,Y),two(X,Y)\cons{}post(X,Y)\cons{}1/post(X,Y),fail\cons{}\nil\}}{\{\lemaatom{X\myordeq{}1}{one(X,Y)}\cons{}\substshow{\bind{X}{1}}\cons{}\nil\}}
 \desno \lineventshow{exit}{Y\myordeq{}a}{\{(1/(Y\myordeq{}a);Y\myordeq{}b)\cons{}two(1,Y)\cons{}2/one(X,Y),two(X,Y)\cons{}post(X,Y)\cons{}1/post(X,Y),fail\cons{}\nil\}}{\{\substshow{\bind{Y}{a}}\cons{}\lemaatom{X\myordeq{}1}{one(X,Y)}\cons{}\substshow{\bind{X}{1}}\cons{}\nil\}}
 \desno \lineventshow{exit}{(Y\myordeq{}a;Y\myordeq{}b)}{\{two(1,Y)\cons{}2/one(X,Y),two(X,Y)\cons{}post(X,Y)\cons{}1/post(X,Y),fail\cons{}\nil\}}{\{\lemaor{Y\myordeq{}a}{(1/(Y\myordeq{}a);Y\myordeq{}b)}\cons{}\substshow{\bind{Y}{a}}\cons{}\lemaatom{X\myordeq{}1}{one(X,Y)}\cons{}\substshow{\bind{X}{1}}\cons{}\nil\}}
 \desno \lineventshow{exit}{two(1,Y)}{\{2/one(X,Y),two(X,Y)\cons{}post(X,Y)\cons{}1/post(X,Y),fail\cons{}\nil\}}{\{\lemaatom{(Y\myordeq{}a;Y\myordeq{}b)}{two(1,Y)}\cons{}\lemaor{Y\myordeq{}a}{(1/(Y\myordeq{}a);Y\myordeq{}b)}\cons{}\substshow{\bind{Y}{a}}\cons{}\lemaatom{X\myordeq{}1}{one(X,Y)}\cons{}\substshow{\bind{X}{1}}\cons{}\nil\}}
 \desno \lineventshow{exit}{(one(X,Y),two(X,Y))}{\{post(X,Y)\cons{}1/post(X,Y),fail\cons{}\nil\}}{\{\lemaatom{(Y\myordeq{}a;Y\myordeq{}b)}{two(1,Y)}\cons{}\lemaor{Y\myordeq{}a}{(1/(Y\myordeq{}a);Y\myordeq{}b)}\cons{}\substshow{\bind{Y}{a}}\cons{}\lemaatom{X\myordeq{}1}{one(X,Y)}\cons{}\substshow{\bind{X}{1}}\cons{}\nil\}}
 \desno \lineventshow{exit}{post(X,Y)}{\{1/post(X,Y),fail\cons{}\nil\}}{\{\lemaatom{(one(X,Y),two(X,Y))}{post(X,Y)}\cons{}\lemaatom{(Y\myordeq{}a;Y\myordeq{}b)}{two(1,Y)}\cons{}\lemaor{Y\myordeq{}a}{(1/(Y\myordeq{}a);Y\myordeq{}b)}\cons{}\substshow{\bind{Y}{a}}\cons{}\lemaatom{X\myordeq{}1}{one(X,Y)}\cons{}\substshow{\bind{X}{1}}\cons{}\nil\}}
 \desno \lineventshow{call}{fail}{\{2/post(X,Y),fail\cons{}\nil\}}{\{\lemaatom{(one(X,Y),two(X,Y))}{post(X,Y)}\cons{}\lemaatom{(Y\myordeq{}a;Y\myordeq{}b)}{two(1,Y)}\cons{}\lemaor{Y\myordeq{}a}{(1/(Y\myordeq{}a);Y\myordeq{}b)}\cons{}\substshow{\bind{Y}{a}}\cons{}\lemaatom{X\myordeq{}1}{one(X,Y)}\cons{}\substshow{\bind{X}{1}}\cons{}\nil\}}
 \desno \lineventshow{fail}{fail}{\{2/post(X,Y),fail\cons{}\nil\}}{\{\lemaatom{(one(X,Y),two(X,Y))}{post(X,Y)}\cons{}\lemaatom{(Y\myordeq{}a;Y\myordeq{}b)}{two(1,Y)}\cons{}\lemaor{Y\myordeq{}a}{(1/(Y\myordeq{}a);Y\myordeq{}b)}\cons{}\substshow{\bind{Y}{a}}\cons{}\lemaatom{X\myordeq{}1}{one(X,Y)}\cons{}\substshow{\bind{X}{1}}\cons{}\nil\}}
 \desno \lineventshow{redo}{post(X,Y)}{\{1/post(X,Y),fail\cons{}\nil\}}{\{\lemaatom{(one(X,Y),two(X,Y))}{post(X,Y)}\cons{}\lemaatom{(Y\myordeq{}a;Y\myordeq{}b)}{two(1,Y)}\cons{}\lemaor{Y\myordeq{}a}{(1/(Y\myordeq{}a);Y\myordeq{}b)}\cons{}\substshow{\bind{Y}{a}}\cons{}\lemaatom{X\myordeq{}1}{one(X,Y)}\cons{}\substshow{\bind{X}{1}}\cons{}\nil\}}
 \desno \lineventshow{redo}{(one(X,Y),two(X,Y))}{\{post(X,Y)\cons{}1/post(X,Y),fail\cons{}\nil\}}{\{\lemaatom{(Y\myordeq{}a;Y\myordeq{}b)}{two(1,Y)}\cons{}\lemaor{Y\myordeq{}a}{(1/(Y\myordeq{}a);Y\myordeq{}b)}\cons{}\substshow{\bind{Y}{a}}\cons{}\lemaatom{X\myordeq{}1}{one(X,Y)}\cons{}\substshow{\bind{X}{1}}\cons{}\nil\}}
 \desno \lineventshow{redo}{two(X,Y)}{\{2/one(X,Y),two(X,Y)\cons{}post(X,Y)\cons{}1/post(X,Y),fail\cons{}\nil\}}{\{\lemaatom{(Y\myordeq{}a;Y\myordeq{}b)}{two(1,Y)}\cons{}\lemaor{Y\myordeq{}a}{(1/(Y\myordeq{}a);Y\myordeq{}b)}\cons{}\substshow{\bind{Y}{a}}\cons{}\lemaatom{X\myordeq{}1}{one(X,Y)}\cons{}\substshow{\bind{X}{1}}\cons{}\nil\}}
 \desno \lineventshow{redo}{(Y\myordeq{}a;Y\myordeq{}b)}{\{two(1,Y)\cons{}2/one(X,Y),two(X,Y)\cons{}post(X,Y)\cons{}1/post(X,Y),fail\cons{}\nil\}}{\{\lemaor{Y\myordeq{}a}{(1/(Y\myordeq{}a);Y\myordeq{}b)}\cons{}\substshow{\bind{Y}{a}}\cons{}\lemaatom{X\myordeq{}1}{one(X,Y)}\cons{}\substshow{\bind{X}{1}}\cons{}\nil\}}
 \desno \lineventshow{redo}{Y\myordeq{}a}{\{(1/(Y\myordeq{}a);Y\myordeq{}b)\cons{}two(1,Y)\cons{}2/one(X,Y),two(X,Y)\cons{}post(X,Y)\cons{}1/post(X,Y),fail\cons{}\nil\}}{\{\substshow{\bind{Y}{a}}\cons{}\lemaatom{X\myordeq{}1}{one(X,Y)}\cons{}\substshow{\bind{X}{1}}\cons{}\nil\}}
 \desno \lineventshow{fail}{Y\myordeq{}a}{\{(1/(Y\myordeq{}a);Y\myordeq{}b)\cons{}two(1,Y)\cons{}2/one(X,Y),two(X,Y)\cons{}post(X,Y)\cons{}1/post(X,Y),fail\cons{}\nil\}}{\{\lemaatom{X\myordeq{}1}{one(X,Y)}\cons{}\substshow{\bind{X}{1}}\cons{}\nil\}}
 \desno \lineventshow{call}{Y\myordeq{}b}{\{(2/(Y\myordeq{}a);Y\myordeq{}b)\cons{}two(1,Y)\cons{}2/one(X,Y),two(X,Y)\cons{}post(X,Y)\cons{}1/post(X,Y),fail\cons{}\nil\}}{\{\lemaatom{X\myordeq{}1}{one(X,Y)}\cons{}\substshow{\bind{X}{1}}\cons{}\nil\}}
 \desno \lineventshow{exit}{Y\myordeq{}b}{\{(2/(Y\myordeq{}a);Y\myordeq{}b)\cons{}two(1,Y)\cons{}2/one(X,Y),two(X,Y)\cons{}post(X,Y)\cons{}1/post(X,Y),fail\cons{}\nil\}}{\{\substshow{\bind{Y}{b}}\cons{}\lemaatom{X\myordeq{}1}{one(X,Y)}\cons{}\substshow{\bind{X}{1}}\cons{}\nil\}}
 \desno \lineventshow{exit}{(Y\myordeq{}a;Y\myordeq{}b)}{\{two(1,Y)\cons{}2/one(X,Y),two(X,Y)\cons{}post(X,Y)\cons{}1/post(X,Y),fail\cons{}\nil\}}{\{\lemaor{Y\myordeq{}b}{(2/(Y\myordeq{}a);Y\myordeq{}b)}\cons{}\substshow{\bind{Y}{b}}\cons{}\lemaatom{X\myordeq{}1}{one(X,Y)}\cons{}\substshow{\bind{X}{1}}\cons{}\nil\}}
 \desno \lineventshow{exit}{two(1,Y)}{\{2/one(X,Y),two(X,Y)\cons{}post(X,Y)\cons{}1/post(X,Y),fail\cons{}\nil\}}{\{\lemaatom{(Y\myordeq{}a;Y\myordeq{}b)}{two(1,Y)}\cons{}\lemaor{Y\myordeq{}b}{(2/(Y\myordeq{}a);Y\myordeq{}b)}\cons{}\substshow{\bind{Y}{b}}\cons{}\lemaatom{X\myordeq{}1}{one(X,Y)}\cons{}\substshow{\bind{X}{1}}\cons{}\nil\}}
 \desno \lineventshow{exit}{(one(X,Y),two(X,Y))}{\{post(X,Y)\cons{}1/post(X,Y),fail\cons{}\nil\}}{\{\lemaatom{(Y\myordeq{}a;Y\myordeq{}b)}{two(1,Y)}\cons{}\lemaor{Y\myordeq{}b}{(2/(Y\myordeq{}a);Y\myordeq{}b)}\cons{}\substshow{\bind{Y}{b}}\cons{}\lemaatom{X\myordeq{}1}{one(X,Y)}\cons{}\substshow{\bind{X}{1}}\cons{}\nil\}}
 \desno \lineventshow{exit}{post(X,Y)}{\{1/post(X,Y),fail\cons{}\nil\}}{\{\lemaatom{(one(X,Y),two(X,Y))}{post(X,Y)}\cons{}\lemaatom{(Y\myordeq{}a;Y\myordeq{}b)}{two(1,Y)}\cons{}\lemaor{Y\myordeq{}b}{(2/(Y\myordeq{}a);Y\myordeq{}b)}\cons{}\substshow{\bind{Y}{b}}\cons{}\lemaatom{X\myordeq{}1}{one(X,Y)}\cons{}\substshow{\bind{X}{1}}\cons{}\nil\}}
 \desno \lineventshow{call}{fail}{\{2/post(X,Y),fail\cons{}\nil\}}{\{\lemaatom{(one(X,Y),two(X,Y))}{post(X,Y)}\cons{}\lemaatom{(Y\myordeq{}a;Y\myordeq{}b)}{two(1,Y)}\cons{}\lemaor{Y\myordeq{}b}{(2/(Y\myordeq{}a);Y\myordeq{}b)}\cons{}\substshow{\bind{Y}{b}}\cons{}\lemaatom{X\myordeq{}1}{one(X,Y)}\cons{}\substshow{\bind{X}{1}}\cons{}\nil\}}
 \desno \lineventshow{fail}{fail}{\{2/post(X,Y),fail\cons{}\nil\}}{\{\lemaatom{(one(X,Y),two(X,Y))}{post(X,Y)}\cons{}\lemaatom{(Y\myordeq{}a;Y\myordeq{}b)}{two(1,Y)}\cons{}\lemaor{Y\myordeq{}b}{(2/(Y\myordeq{}a);Y\myordeq{}b)}\cons{}\substshow{\bind{Y}{b}}\cons{}\lemaatom{X\myordeq{}1}{one(X,Y)}\cons{}\substshow{\bind{X}{1}}\cons{}\nil\}}
 \desno \lineventshow{redo}{post(X,Y)}{\{1/post(X,Y),fail\cons{}\nil\}}{\{\lemaatom{(one(X,Y),two(X,Y))}{post(X,Y)}\cons{}\lemaatom{(Y\myordeq{}a;Y\myordeq{}b)}{two(1,Y)}\cons{}\lemaor{Y\myordeq{}b}{(2/(Y\myordeq{}a);Y\myordeq{}b)}\cons{}\substshow{\bind{Y}{b}}\cons{}\lemaatom{X\myordeq{}1}{one(X,Y)}\cons{}\substshow{\bind{X}{1}}\cons{}\nil\}}
 \desno \lineventshow{redo}{(one(X,Y),two(X,Y))}{\{post(X,Y)\cons{}1/post(X,Y),fail\cons{}\nil\}}{\{\lemaatom{(Y\myordeq{}a;Y\myordeq{}b)}{two(1,Y)}\cons{}\lemaor{Y\myordeq{}b}{(2/(Y\myordeq{}a);Y\myordeq{}b)}\cons{}\substshow{\bind{Y}{b}}\cons{}\lemaatom{X\myordeq{}1}{one(X,Y)}\cons{}\substshow{\bind{X}{1}}\cons{}\nil\}}
 \desno \lineventshow{redo}{two(X,Y)}{\{2/one(X,Y),two(X,Y)\cons{}post(X,Y)\cons{}1/post(X,Y),fail\cons{}\nil\}}{\{\lemaatom{(Y\myordeq{}a;Y\myordeq{}b)}{two(1,Y)}\cons{}\lemaor{Y\myordeq{}b}{(2/(Y\myordeq{}a);Y\myordeq{}b)}\cons{}\substshow{\bind{Y}{b}}\cons{}\lemaatom{X\myordeq{}1}{one(X,Y)}\cons{}\substshow{\bind{X}{1}}\cons{}\nil\}}
 \desno \lineventshow{redo}{(Y\myordeq{}a;Y\myordeq{}b)}{\{two(1,Y)\cons{}2/one(X,Y),two(X,Y)\cons{}post(X,Y)\cons{}1/post(X,Y),fail\cons{}\nil\}}{\{\lemaor{Y\myordeq{}b}{(2/(Y\myordeq{}a);Y\myordeq{}b)}\cons{}\substshow{\bind{Y}{b}}\cons{}\lemaatom{X\myordeq{}1}{one(X,Y)}\cons{}\substshow{\bind{X}{1}}\cons{}\nil\}}
 \desno \lineventshow{redo}{Y\myordeq{}b}{\{(2/(Y\myordeq{}a);Y\myordeq{}b)\cons{}two(1,Y)\cons{}2/one(X,Y),two(X,Y)\cons{}post(X,Y)\cons{}1/post(X,Y),fail\cons{}\nil\}}{\{\substshow{\bind{Y}{b}}\cons{}\lemaatom{X\myordeq{}1}{one(X,Y)}\cons{}\substshow{\bind{X}{1}}\cons{}\nil\}}
 \desno \lineventshow{fail}{Y\myordeq{}b}{\{(2/(Y\myordeq{}a);Y\myordeq{}b)\cons{}two(1,Y)\cons{}2/one(X,Y),two(X,Y)\cons{}post(X,Y)\cons{}1/post(X,Y),fail\cons{}\nil\}}{\{\lemaatom{X\myordeq{}1}{one(X,Y)}\cons{}\substshow{\bind{X}{1}}\cons{}\nil\}}
 \desno \lineventshow{fail}{(Y\myordeq{}a;Y\myordeq{}b)}{\{two(1,Y)\cons{}2/one(X,Y),two(X,Y)\cons{}post(X,Y)\cons{}1/post(X,Y),fail\cons{}\nil\}}{\{\lemaatom{X\myordeq{}1}{one(X,Y)}\cons{}\substshow{\bind{X}{1}}\cons{}\nil\}}
 \desno \lineventshow{fail}{two(1,Y)}{\{2/one(X,Y),two(X,Y)\cons{}post(X,Y)\cons{}1/post(X,Y),fail\cons{}\nil\}}{\{\lemaatom{X\myordeq{}1}{one(X,Y)}\cons{}\substshow{\bind{X}{1}}\cons{}\nil\}}
 \desno \lineventshow{redo}{one(X,Y)}{\{1/one(X,Y),two(X,Y)\cons{}post(X,Y)\cons{}1/post(X,Y),fail\cons{}\nil\}}{\{\lemaatom{X\myordeq{}1}{one(X,Y)}\cons{}\substshow{\bind{X}{1}}\cons{}\nil\}}
 \desno \lineventshow{redo}{X\myordeq{}1}{\{one(X,Y)\cons{}1/one(X,Y),two(X,Y)\cons{}post(X,Y)\cons{}1/post(X,Y),fail\cons{}\nil\}}{\{\substshow{\bind{X}{1}}\cons{}\nil\}}
 \desno \lineventshow{fail}{X\myordeq{}1}{\{one(X,Y)\cons{}1/one(X,Y),two(X,Y)\cons{}post(X,Y)\cons{}1/post(X,Y),fail\cons{}\nil\}}{\{\nil\}}
 \desno \lineventshow{fail}{one(X,Y)}{\{1/one(X,Y),two(X,Y)\cons{}post(X,Y)\cons{}1/post(X,Y),fail\cons{}\nil\}}{\{\nil\}}
 \desno \lineventshow{fail}{(one(X,Y),two(X,Y))}{\{post(X,Y)\cons{}1/post(X,Y),fail\cons{}\nil\}}{\{\nil\}}
 \desno \lineventshow{fail}{post(X,Y)}{\{1/post(X,Y),fail\cons{}\nil\}}{\{\nil\}}
 \desno \lineventshow{fail}{(post(X,Y),fail)}{\{\nil\}}{\{\nil\}}
\end{myv0}
\caption{Execution of a query in \mymodel}
\end{mytable0}
\addtolength{\textwidth}{-\deltawid} 
\end{landscape}
\end{appendix}

\end{document}